\documentclass{PoS}

\title{A search for H$^{+}$ and H$^{++}$ Higgs boson with the CMS detector}

\ShortTitle{A search for H$^{+}$ and H$^{++}$ Higgs boson with the CMS detector}

\author{\speaker{Aruna Kumar NAYAK}%
        \thanks{On behalf of CMS Collaboration}\\
       LIP-Laboratory for Instrumentation and Experimental Particle Physics, Lisbon \\
       E-mail: \email{Aruna.Nayak@cern.ch}}

\abstract{We present results from a search for an exotic Higgs boson in the channel H$^{++}$ $\rightarrow ~\ell^{+}\ell^{+}$ with the CMS detector using data accumulated in the 2010 \& 2011 running of the LHC at $\sqrt{s}$ = 7 TeV.  We also present results from a search for a charged Higgs boson in $t\bar{t}$ decays in the channel H$^{+}$ $\rightarrow$ $\tau\nu$.
}

\FullConference{XXIst International Europhysics Conference on High Energy Physics\\
                21-27 July 2011\\
                Grenoble, Rhones Alpes France}

\begin{document}

\section{Introduction}
In the Minimal Supersymmetric extension to the Standard Model (MSSM) five Higgs bosons are predicted: 
the CP-even h$^{0}$ and H$^{0}$, the pseudoscalar A$^{0}$ and the charged Higgs boson H$^{+}$~\cite{PAS_HIG_11_008}.
If the charged Higgs boson mass is smaller than the top quark mass, i.e. $m_{H^{+}} < m_{t} - m_b$, 
the top quark can decay via 
$t\rightarrow H^{+} b$ (and its charge conjugate). 
For values of $\tan\beta$ (the ratio of the vacuum expectation values of the two Higgs boson doublets) 
larger than 20, the charged 
Higgs boson preferentially decays to $\tau$ lepton and neutrino, $H^{+} \rightarrow \tau ^{+} \nu _{\tau}$. 
The presence of the $t\rightarrow H^{+} b$, $H^{+} \rightarrow \tau ^{+} \nu_{\tau}$ decay modes alters the SM prediction of 
the $\tau$ lepton yield in the decay products of the $t\bar{t}$ pairs.
This article presents the results of the search for a charged Higgs boson in the decay of top quarks using 1.1 fb$^{-1}$ of data accumulated using CMS detector~\cite{JINST:3:S08004}. 

Doubly-charged Higgs bosons ($\Phi^{++}$) appear in exotic Higgs representations such as found in minimal seesaw model of type II and left-right symmetric models~\cite{PAS_HIG_11_007}.
This particle decays to the same charged lepton pairs $\ell^+_i\ell^+_j$ allowing lepton flavor violating decays. 
In the see-saw mechanism of type-II the  $\Phi^{++}$ Yukawa coupling matrix $Y_\Phi^{ij}$ is proportional to the light neutrino mass matrix and allows to test the neutrino mass mechanism by measuring the  branching fractions $\Phi^{++}\to \ell_i\ell_j$ at the LHC. The results of an inclusive search for the doubly charged Higgs boson is also presented.

\section{Search for charged Higgs boson in decay of top quarks}
The dominant source of top quarks at LHC is $pp \rightarrow t\bar{t}$ process, therefore the charged Higgs boson has been searched for 
in the subsequent decay products of the top quark pairs: $t \bar{t} \rightarrow H^{\pm} W^{\mp} b \bar{b}$ and 
$t \bar{t} \rightarrow H^{\pm} H^{\mp} b \bar{b}$ when $H^{\pm}$ decays into $\tau$ lepton and neutrino~\cite{PAS_HIG_11_008}. 
Three types of the final  states with large missing transverse energy and b-tagged jets have been analyzed: with e+$\mu$ pair, with muon plus $\tau$ 
decaying hadronically ($\mu \tau_{h}$) and with no leptons and $\tau$ decaying hadronically (fully hadronic final state). 

In the e$\mu$ channel the events are triggered by the combined electron plus muon triggers.
In the offline analysis the events are selected by requiring at least one isolated electron
and one isolated muon with $E_{T}>$ 20 GeV ($p_{T}>$ 20 GeV/$c$), $| \eta |<$ 2.5 (2.4) for electrons (muons),
at least two jets with $p_{T}>$30 GeV/$c$, $|\eta|<$2.4.
The jets are required to be separated from the leptons with $\Delta R>$0.4.
In the dileptonic $\mu\tau$ channel, the events are triggered by the single muon triggers. 
In the offline analysis events are selected by requiring one isolated, high $p_{T}$ muon,
$p_{T}>$ 20 GeV/$c$ and $|\eta|<$2.1, at least two jets with $p_{T}>$30 GeV/$c$ and $|\eta
|<$2.4 with at least one b-tagged jet,
the missing $E_{T}>$ 40 GeV and one $\tau_{h}$ with $p_{T}>$ 20 GeV/$c$ and $|\eta|<2.4$. 
In the fully-hadronic channel the events are triggered by the combined $\tau$ plus $E_{T}^{miss}$ trigger. 
In the off-line analysis the event selection includes the requirements to have one $\tau_{h}$
with $p_{T}^{\tau_{h}}>$ 40 GeV/$c$, $| \eta | <$ 2.1,
at least three other jets ($N_{j} \geq$ 3) with $p_{T}>$ 30 GeV/$c$ and
$| \eta |<$ 2.4 with at least one b-tagged jet, no isolated leptons (muons or electrons) with $p_{T}>$ 15 GeV/$c$
and $| \eta | <$ 2.5 (lepton veto), the $E_{T}^{miss}>$70 GeV.
In order to effectively suppress the QCD multi-jet background only one prong $\tau_{h}$ decays are selected with the additional cut on the transverse momentum of the leading charged particle, $p_{T}^{trk}>$ 20 GeV/$c$.

The backgrounds considered are the QCD multi-jets production,
$W$+jets and $Z$+jets processes, di-boson ($WW$, $ZZ$, $WZ$) production, the SM $t \bar{t}$ and single top production dominated by $tW$ channel. 
The major backgrounds in fully-hadronic channel are the electroweak and $t \bar{t}$ backgrounds containing a real hadronic tau and the QCD multi-jets background. These are estimated using data. 
The remaining electroweak and $t \bar{t}$ backgrounds, where an electron, muon or a jet is mis-identified as a tau jet, are estimated using Monte Carlo (MC) simulated events. 
The major backgrounds in the $\mu\tau$ channel are the SM $t \bar{t}$ and $W$+jets.
The $W$+jets, $t \bar{t} \rightarrow \ell+jets$ and QCD multijets backgrounds, where a jet is mis-identified as a hadronic decaying tau, are estimated using data. 
The other backgrounds containing a real tau are estimated from MC simulated events. 
The SM $t \bar{t}$ process constitutes the largest part of the backgrounds in the leptonic e$\mu$ channel. All the backgrounds are estimated using MC simulated events.  

A good agreement is found between the number of observed events from data and the expected SM backgrounds within the uncertainties. No excess (or lack) of events is observed in any channel. 

\begin{figure}[htp]
\begin{center}
\begin{tabular}{c}
\includegraphics[width=0.45\textwidth]{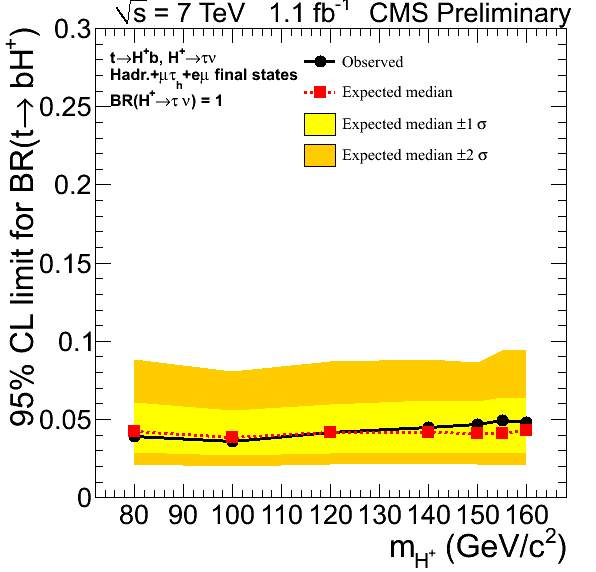} \hfill
\includegraphics[width=0.45\textwidth]{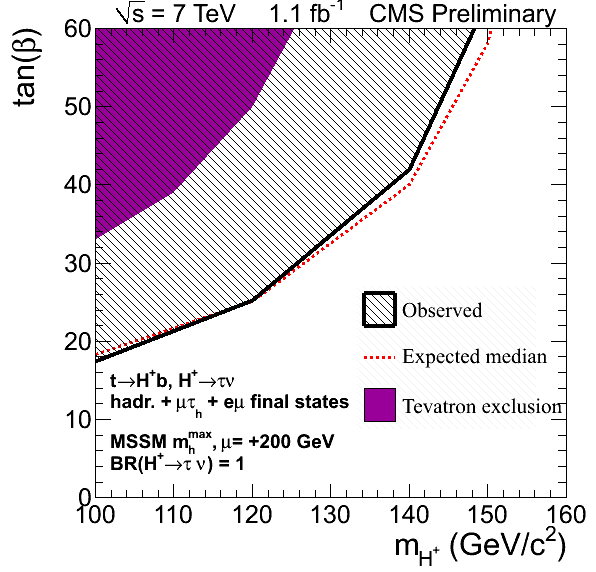} 
\end{tabular}
\caption{Left : Upper limit on $BR(t\rightarrow H^{+}b)$ assuming $BR(H^{+} \rightarrow \tau \nu)$=1 as a function of $m_{H^{+}}$
for the combination of the fully hadronic, the $\mu\tau_{h}$ and the e$\mu$ final states. 
The yellow bands show the one and two $\sigma$ uncertainties around the expected limit.
Right : The exclusion region in the MSSM $M_{H^{+}}$-tan$\beta$ parameter space obtained from the combined analysis for the MSSM $m_{h}^{max}$ scenario. The exclusion region obtained by the Tevatron experiments~\cite{PAS_HIG_11_008} are also shown.
}
\label{fig:HplusLimit}
\end{center}
\end{figure}

A CLs method is used in order to obtain the upper limit at 95\%~C.L. on branching ratio $BR(t \rightarrow bH^{+})$~\cite{PAS_HIG_11_008}. 
Figure~\ref{fig:HplusLimit} (left) shows the upper limit on the branching ratio $BR(t\rightarrow H^{+}b)$ assuming $BR(H^{+} \rightarrow \tau \nu)$=1 as a function of $m_{H^{+}}$ obtained from combining all three final states. 
Figure~\ref{fig:HplusLimit} (right) shows the exclusion region in the MSSM $M_{H^{+}}$-tan$\beta$ parameter space for the MSSM $m_{h}^{max}$ scenario~\cite{PAS_HIG_11_008}. 

\section{Search for doubly charged Higgs boson}
At the LHC, the doubly charged Higgs boson can be produced via both the pair production process $pp \to \Phi^{++}\Phi^{--}\to \ell_i^+\ell_j^+\ell_k^-\ell_l^-$
as well as  the associated production process $pp \rightarrow \Phi^{++}\Phi^{-}\to  \ell_i^+\ell_j^+\ell_k^-\nu_l$~\cite{PAS_HIG_11_007}. 
Both production processes are studied, assuming that the $\Phi^{++}$ and $\Phi^{+}$ are degenerate in mass. 
We search for an excess of events in all possible flavour combinations of the same charge
lepton pairs coming from the decays $\Phi^{++}\to \ell^+_i\ell^+_j$ without making
assumptions on the $\Phi^{++}$ branching fractions.
Both the three and four charged lepton final states are considered including at most one and two $\tau$ leptons, respectively.
In our search the $\Phi^{++}\to W^{+}W^{+}$ decays are assumed to be suppressed.
In addition to the model independent search, the type II seesaw model is tested in four benchmark points (BP)~\cite{PAS_HIG_11_007} that characterize different characteristic neutrino mass matrix structures. 

The most important signature of the $\Phi^{++}$ search is the presence of two resonant same charge leptons. The occurrence of such a signature is extremely rare in SM processes.
For the four lepton final state from $\Phi^{++}\Phi^{--}$ pair production both doubly charged Higgs bosons could be reconstructed giving two same charge pairs of leptons.

The events are selected at the trigger level using double lepton (ee, e$\mu$, $\mu\mu$) trigger. 
In offline, the events are selected having one or more pairs of same sign dileptons. Additional topological cuts are applied depending on final states as discussed in ref.~\cite{PAS_HIG_11_007}.  
The selection cuts are optimized for significance across various final states and masses.
The full selection for event counting is based on a mass window for a given $m(\Phi)$. The choice of mass window is motivated 
by the requirement of high efficiency for signal events across a variety of final states.
In case of four lepton final states the mass window is defined as a two dimensional region in the plane
of $m(\Phi^{++})$ vs $m(\Phi^{--})$. 

\begin{figure}[htp]
\begin{center}
\begin{tabular}{c}
\includegraphics[width=0.5\textwidth]{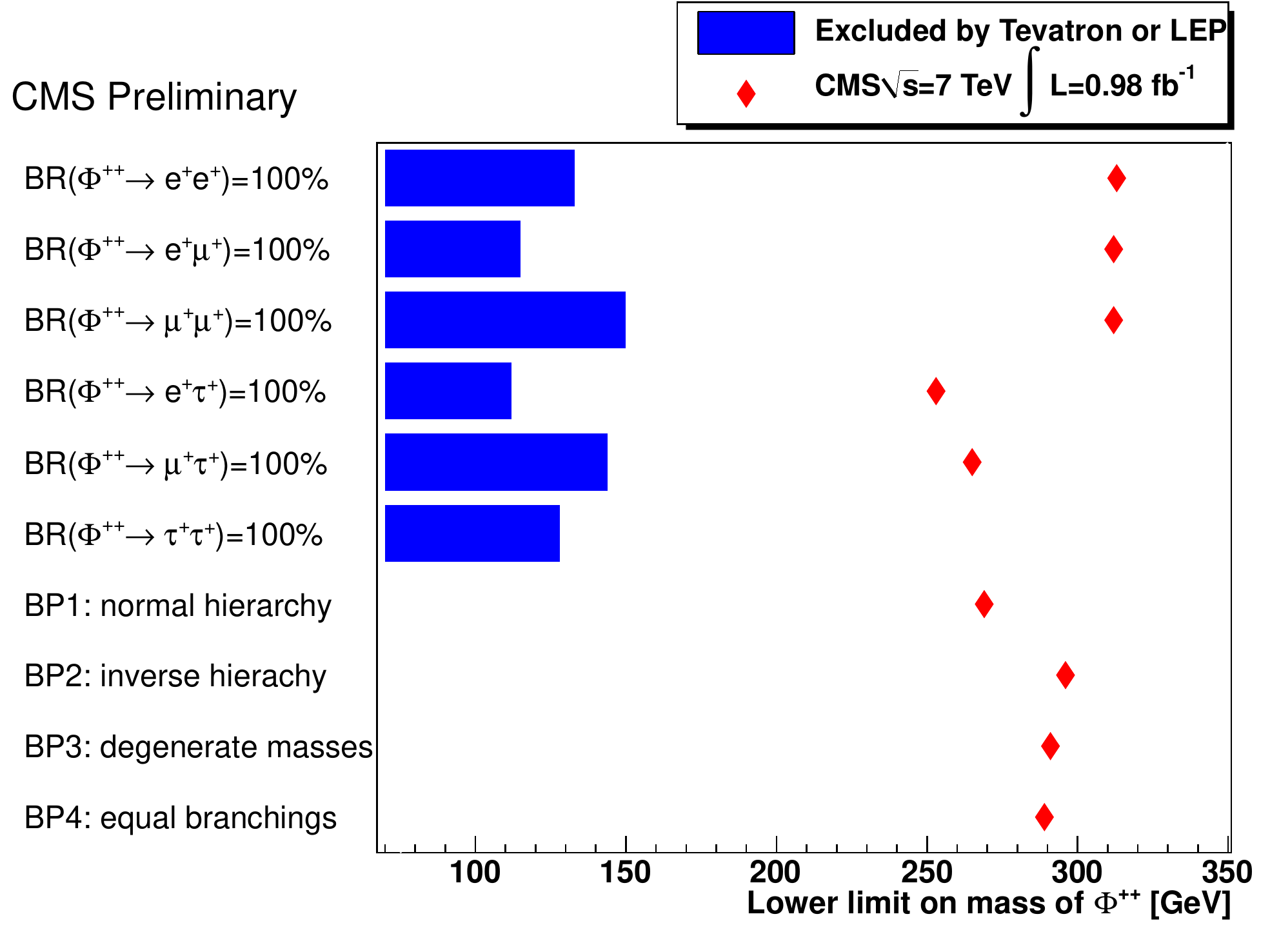} 
\end{tabular}
\caption{Observed lower limit on mass of $\Phi^{++}$ at in 95\% C.L. in different lepton final states. The branching ratio that are assumed in the limit calculation are indicated.}
\label{fig:PhippLimit}
\end{center}
\end{figure}

A good agreement is found between data and the Monte Carlo simulation in all final states. Model independent lower bound on the mass of $\Phi^{++}$ is derived using CLs method. The results of the exclusion limit are shown in Figure~\ref{fig:PhippLimit}. These are the best limits to date in all decay channels except for 100\% decays to $\tau$'s.

\end{document}